\documentclass[12pt]{article}
\usepackage{epsf}
\usepackage{amsmath}

\usepackage{cite}

\usepackage{graphicx}

\begin{document}

\hfill January2026

\noindent
{\bf \LARGE Model of Dark Matter and Energy\\

\vspace{1.5cm}
\noindent
{\bf  \large Paul H. Frampton}\footnote{paul.h.frampton@gmail.com}\\

\vspace{0.1in}

\noindent
{\it \large  Dipartimento di Matematica e Fisica "Ennio De Giorgi",\\ 
Universit\`{a} del Salento and INFN-Lecce,\\ Via Arnesano, 73100 Lecce, Italy.}\\

\vspace{0.5in}

\noindent
{\large Talk at the South African Gravity Society, November 20, 2025.}
{\large SAGS2025, Stonehenge River Lodge, Parys, South Africa, to appear in Proceeding of SAGS2025 Conference, Nov. 18-21, 2025.

\newpage

\section{Introduction}

\bigskip

\noindent
A well-known question in theoretical cosmology concerns the cosmological  constant: why 
does it have the value observed?

\bigskip

\noindent
This constant, $\Lambda$, has the dimensions of a density and observations suggest
that its value is of order 

\begin{equation}
\Lambda = + ~ O\left((meV)^4\right).
\label{CC1}
\end{equation}

\noindent
It is sometimes stated that this value is 
surprisingly small, and ingenious explanations have been offered
for its alleged smallness. 

\bigskip

\noindent
Conventional wisdom is that the energy make-up of the universe is approximately
5\% normal matter, 25\% dark matter and 70\% dark energy where dark energy is
responsible for the observed accelerated expansion.

\bigskip

\noindent 
In the present talk, we discuss a model in which both of these issues are addressed
in a novel way. The cosmological constant naturally appears with a value
consistent with Eq.(\ref{CC1})
without fine-tuning. Our calculations can all be carried out with the Planck
constant $\hbar$. and hence the Planck mass $M_{Planck} \equiv \sqrt{\hbar G/c}$,
set equal to zero. In other words, Eq.(\ref{CC1})
has no connection with quantum mechanics.

\bigskip

\noindent
We shall begin
by providing a very broad-brush discussion of our model.

\bigskip

\noindent 
Some of our later results may seem at first sight
to be counterintuitive.

\bigskip

\noindent
We know of no  fatal flaw.

\bigskip

\noindent
In the model there are four (not three) different ranges of length scale
for each of which the ruling forces are different. At the largest scale (the
universe) it is electromagnetic; at the second scale (clusters, galaxies,
planetary systems) it is gravitational; at the third scale (molecules, atoms)
it is electromagnetic again;  only at the fourth and smallest scale (nucleons, quarks and leptons)
is electromagnetism joined by strong and weak interactions. Note that the first and third 
scales, dominated by electromagnetism 
"sandwich" the second scale, dominated by gravitation. 
\noindent
It seems conceivable that careful study of the consistency of this sandwich
might shed light on how to make a theory of gravity beyond Einstein. 

\bigskip

\noindent
As for the energy make up, in our model
it is 5\% normal matter and 95\% dark matter, with 0\% dark energy. 

\bigskip

\noindent
Dark energy
is replaced by a 70\% part
in the 95\% dark matter, a part which is composed of electrically-charged, 
extremely-massive \\primordial naked singularities (PEMNSs).

\bigskip

 \noindent
\section{ The visible universe}

\bigskip

\noindent
In this section, we establish the mean mass density of
our model universe.

\bigskip

\noindent
Its present co-moving radius is 14 Gpc which translates into a volume

\begin{equation}
V_U \sim 4 \times 10^{89} cm^3 \sim \left( 5 \times 10^{91} (meV)^3 \right)^{-1}
\label{volumeU}
\end{equation}

\noindent
We shall adopt as the mass $M_U$ of the model univer
\begin{equation}
M_U \equiv 10^{23} M_{\odot} \sim 2 \times 10^{92} meV
\end{equation}

\noindent
where we have used the approximation $M_{\odot} = 2 \times 10^{30}$ kg.

\bigskip

\noindent
The density $\rho_U$ in the model is therefore

\begin{equation}
\rho_U = \left( \frac{M_U}{V_U} \right) \sim  (\sqrt{2} meV)^4
\label{density}
\end{equation}
which, by comparison with Eq.(\ref{CC1}), shows that

\begin{equation}
\rho_U = O(\Lambda)
\label{OoM}
\end{equation}
meaning that the mean mass density of our model universe
is within one order of magnitude of the cosmological constant
observed in the real universe.

\bigskip

\noindent
We note that the above values of $M_U$ and
$\rho_U$ are for inclusion only of Old Dark Matter. Both
will be exactly doubled when we add New Dark Matter,
but the order of magnitude statement in Eq.(\ref{OoM}),
the only thing actually used in all of our ensuing 
discussion, will remain valid.

\bigskip

\noindent
\section{ Old dark matter}

\bigskip
\noindent
For the dark matter in galaxies and clusters, we assume the
dark matter constituents are PIMBHs (IM=Intermediate Mass)
with masses in the range

\begin{equation}
100 M_{\odot} < M_{PIMBH} < 10^5 M_{\odot}
\label{Mpimbh}
\end{equation}

\bigskip

\noindent
This assumption could be tested using microlensing of the stars in the
Magellanic Clouds, for which a precursor is the MACHO Collaboration. 
They discovered light curves
corresponding to PIMBH masses only up to $\sim 10 M_{\odot}$.
Checking the mass range in Eq.(\ref{Mpimbh}) cannot be done
quickly as the light curve duration is $\sim 2$ years for
$M_{PIMBH}=100 M_{\odot}$ and increases to $\sim 60$ years
for $M_{PIMBH}=10^5 M_{\odot}$.

\bigskip

\noindent
In our specific model of Old Dark Matter, we take $10^{21}$
PIMBHs, each with mass $100 M_{\odot}$.

\bigskip

\noindent
We include as Old Dark Matter also the supermassive black holes
observed in galactic centres. For these we assume a mass range

\begin{equation}
10^6 M_{\odot} < M_{PSMBH} <  10^{11} M_{\odot}
\label{Mpsmbh}
\end{equation}
and for these, we assume they all are primordial on the basis that 
there seems to be
insufficient time for stellar-collapse black holes to reach
the mass range in Eq.(\ref{Mpsmbh}) by accretion and
merging.

\bigskip

\noindent
In the model we take, for simplicity, PSMBHs all with mass $10^7 M_{\odot}$,
and one in each of the $10^{11}$ galaxies. In other words, the total
number $n_{PSMBH}$ of PSMBHs in the visible universe is
$n_{PSMBH} =10^{11}$.
      
\bigskip

\section{New dark matter $\equiv$ Dark energy}

\bigskip

\noindent
We follow the suggestion made in 2022 that there
exist a number of PEMNSs (EM=Extremely Massive) where

\begin{equation}
10^{12} M_{\odot} < M_{PEMNS} < 10^{22} M_{\odot}
\label{MassPEMBHs}
\end{equation}

\noindent
We shall take for definiteness $M_{PEMNS} = 10^{12} M_{\odot}$
and thus, according to the semi-empirical rule enunciated in 2022, they
each carry negative charge $Q_{PEMNS} \simeq - 4 \times 10^{32}$ Coulombs.
We shall take a number of PEMNSs

\begin{equation}
n_{PEMNS}  = 10^{11}.
\label{Y}
\end{equation}
Their total mass is therefore
$M_{total}(PSMNS) = 10^{23} M_{\odot}$.

\bigskip

\noindent
We are neglecting spin, so each charged black hole is 
described by a Reissner-N\"{o}rdstrom (RN) metric

\begin{equation}
ds^2 = f(r) dt^2 - f(r)^{-1}dr^2 - r^2 d\theta^2 -r^2 \sin^2 \theta d\phi^2
\label{RNmetric}
\end{equation}
where

\begin{equation}
f(r) \equiv \left( 1 - \frac{r_S}{r} + \frac{r_Q}{r^2} \right).
\label{f(r)}
\end{equation}
with

\begin{equation}
r_S =   2 G M ~~~~~~ r_Q= Q^2 G
\label{rSrQ}
\end{equation}

\bigskip

\noindent
The horizon(s) of the RN metric are where

\begin{equation}
f(r) =0
\end{equation}
which gives

\begin{equation}
r_{\pm} = \frac{1}{2} \left( r_S \pm \sqrt{r_S^2 - 4 r_Q^2} \right)
\label{rpm}
\end{equation}

\bigskip

\noindent
For $2r_Q < r_S$, $Q < M$, there are two horizons. When $2r_Q = r_S$,
$Q=M$ 
the RN black hole is extremal and there is only one horizon.

\bigskip

\noindent
If $2r_Q > r_S$,  $Q > M$, the  RN black hole is super-extremal, there is no horizon
at all  and the $r=0$ singularity is observable to a distant observer. This is known as a naked
singularity. 

\bigskip

\noindent
All of the PEMNSs are super-extremal RN black holes so our model universe
contains precisely a hundred billion naked singularities. We are aware of the
cosmic censorship hypothesis which would allow only zero
naked singularities but that hypothesis is, to our knowledge, unproven so we believe
this is not a fatal flaw.

\bigskip

\noindent
The suggestion in 2022, which the present talk will support,
is that the Coulomb repulsion between PEMNSs could explain the observed
accelerating expansion of the universe. If so, it must lead to a negative
pressure as one requirement and we shall show that this actually occurs.

\bigskip

\noindent
More importantly, we shall show that the
magnitude of this pressure is consistent with the observed equation of state
associated with the cosmological constant.

\bigskip

\noindent
Given the size of the visible universe discussed in the introduction,
it is straightforward to estimate that the mean separation 
$\bar{L}$ of the PEMNSs is a few parsecs while
their Schwarzschild radius is $r_S \sim 0.1$ pc, The wide separation,
with $r_S/\bar{L} \sim 10^{-7}$, implies that an expansion in $1/r$ is
rapidly convergent and this fact will simplify our derivation of the
pressure.

\bigskip

\noindent
To derive the pressure we shall need to evaluate the gravitational
stress-energy pseudotensor

\begin{eqnarray}
T_{\mu\nu}^{(GRAV)} &=& -\frac{1}{8\pi G} \left( G_{\mu\nu} + \Lambda g_{\mu\nu} \right) + \nonumber \\
& &\frac{1}{16 \pi G (-g)} \left( (-g)(g_{\mu\nu}g_{\alpha\beta}-g_{\mu\alpha}g_{\nu\beta}) \right)_{,\alpha\beta} \nonumber 
\label{TGRAV}
\end{eqnarray}

\noindent
where the final subscripts represent simple partial derivatives. We shall need also to evaluate the electromagnetic
counterpart

\begin{equation}
T_{\mu\nu}^{(EM)}  =  F_{\mu\alpha} g^{\alpha\beta} F_{\mu\beta} 
- \frac{1}{4} g_{\mu\nu} F^{\alpha\beta}F_{\alpha\beta}.
\label{TEM}
\end{equation}

\bigskip

\noindent
Let us begin with $T_{\mu\nu}^{(GRAV)}$. This calculation involves taking up to
the second derivatives of the metric. 

\bigskip

\noindent
For the ubiquitous
function $f(r)$:

\begin{equation}
\frac{\partial}{\partial r} f(r) \sim O(1/r^2)  ~~~~~ \frac{\partial^2}{\partial^2 r} f(r) \sim O(1/r^4)
\label{f1}
\end{equation}
\begin{equation}
\frac{\partial}{\partial r} (f(r))^2 \sim O(1/r^2)  ~~~~~ \frac{\partial^2}{\partial^2 r} (f(r))^2 \sim O(1/r^4)
\label{f2}
\end{equation}
\begin{equation}
\frac{\partial}{\partial r} (f(r))^{-1} \sim O(1/r^2)  ~~~~~ \frac{\partial^2}{\partial^2 r} (f(r))^{-1} \sim O(1/r^4)
\label{f3}
\end{equation}
\begin{equation}
\frac{\partial}{\partial r} (f(r))^{-2} \sim O(1/r^2)  ~~~~~ \frac{\partial^2}{\partial^2 r} (f(r))^{-2} \sim O(1/r^4)
\label{f4}
\end{equation}

\bigskip

\noindent
It is not difficult to check, using these derivatives, that the 1st, 3rd and 4th terms
in Eq.(\ref{TGRAV}) all fall off as $O(1/r^2)$ and that only the 2nd term does not.

\bigskip

\noindent
Turning to $T_{\mu\nu}^{(EM)}$, and using the fact that the gauge potential
for an RN naked singilarity is

\begin{equation}
A_{\mu} = \left( \frac{Q}{r}, 0, 0, 0 \right)
\label{Amu}
\end{equation}

\noindent
we can straightforwardly see that both terms in Eq.(\ref{TEM}) fall off as $O(1/r^2)$.

\bigskip

\noindent
Collecting results for the two pieces, gravitational and electromagnetic, of the stress-energy
tensor we deduce that

\begin{equation}
T_{\mu\nu}^{(GRAV)} + T_{\mu\nu}^{(EM)} = - \left( \frac{\Lambda}{8\pi G} \right) g_{\mu\nu} + O(1/r^2).
\label{Ttotal}
\end{equation}

\bigskip

\noindent
The cosmological constant is predicted to be

\begin{equation}
\Lambda \sim  + ~ O\left((meV)^4\right).
\label{CC}
\end{equation}
which is consistent with its observed value.

\bigskip

\noindent
Since the stress-energy tensor is proportional to the metric tensor, and bearing in mind the rapid convergence
of the $(1/r)$ expansion, the equation of state
is predicted to be

\begin{equation}
\omega = P/\rho = -1 \pm O(10^{-14})
\end{equation}
extremely close to the value when the cosmological term in the
generalised Einstein tensor is proportional to the metric.

\bigskip

\noindent
Assuming an FLRW metric for the visible universe, and zero curvature, the Friedmann equation is,
ignoring radiation,

\begin{equation}
\left( \frac{\dot{a}}{a} \right)^2= \frac{\Lambda}{3} + \frac {8\pi}{3} \rho_{matter}
\end{equation}
so that the PEMNSs of the New Dark Matter provide a cosmological component
indistinguishable, as far as the expansion properties of the visible universe are concerned, from
what was called dark energy. Hence: New Dark Matter $\equiv$ Dark Energy.

\bigskip

\noindent
\section{Entropy}

\bigskip

\noindent
Of the known constituents in the universe the entropy is dominated
by the supermassive black holes, PSMBHs at the galactic centres. In our model there are $10^{11}$
galaxies each with a $10^7M_{\odot}$ supermassive black hole at its centre.
Using the well-known PBH entropy formula 
for black holes

\begin{equation}
\left( \frac{S}{k} \right)_{BH} (\eta M_{\odot}) \simeq 10^{78} \eta^2
\label{BHentropy}
\end{equation}

\noindent
the PSMBHs are seen to contribute

\begin{equation}
S/k (PSMBHs) \sim 10^{103}.
\label{S/kPSMBHs}
\end{equation}

\noindent
to the entropy of the universe.

\bigskip

\noindent
The holographic maximum entropy of the universe is its
surface area in units of $L_{Planck}^2 \sim 10^{-66} cm^2$,
namely

\begin{equation}
S/k (holog) \sim \frac {4 \pi (14Gpc)^2}{L_{Planck}^2} \sim 10^{123}
\label{holography}
\end{equation}
so that the PSMBH entropy in Eq.(\ref{S/kPSMBHs}) falls far short
of this, by some twenty orders of magnitude.

\bigskip

\noindent
For the dark matter in galaxies necessary to explain observed
rotation curves, we include PIMBHs. We take $10^{11}$
of them each with mass $M_{PIMBH}=100M_{\odot}$ which leads to

\begin{equation}
S/k (PIMBHs) \sim 10^{103}
\label{S/kPIMBHs}
\end{equation}
which, by coincidence, approximately equals the entropy in Eq.(\ref{S/kPSMBHs})
from PSMBHs.

\bigskip

\noindent
A significant increase towards the holographic bound 
arises when we add the new dark matter, the PEMNSs. If we take
$10^{11}$ PEMNSs each with mass $10^{12} M_{\odot}$ the
entropy is increased by ten orders of magnitude to

\begin{equation}
S/k (PEMNSs) \sim 10^{113}.
\label{S/kPEMBHs}
\end{equation}
which, if the aim is to saturate the holographic limit, is a considerable
improvement.  It cannot be
fully saturated because then the universe itself would be a black hole.

\bigskip

\noindent
If we calculate the Schwarzschild radius of the visible universe
from $R_{VU}^{Schw} \equiv 2 GM_{VU}/c^2$ we find $R_{VU}^{Schw} = 30$Gly
which is 2/3 of the value $R_{VU}=45$Gly. This supports our belief that
there exist in Nature missing components with extremely high entropy.

\bigskip

\noindent
We can achieve saturation by increasing the PEMNS mass to
$10^{12+p} M_{\odot}$ and reducing correspondingly the number to
$n_{PEMNS}=10^{11-p}$ in order to arrive at the revised estimate

\begin{equation}
S/k (PEMNSs) \sim 10^{113+p}
\label{saturation}
\end{equation}

\noindent
so that $p \rightarrow 10$ gives the required additional orders
of magnitude to approach saturation. A naked singularity with mass 
approaching $10^{22} M_{\odot}$ is counterintuitive, but dark matter is so
mysterious that we should not leave any stone unturned.

\bigskip

\noindent
\section{Time Dependence}

\bigskip

\noindent
Our discussion has focused on the present cosmological
time $t=t_0\simeq 13.8$ Gy and already provided some counterintuitive ideas such as that at the largest cosmological distances, {\it e.g.} greater than 1 Gpc, the dominant force
is electromagnetism rather than gravitation. 
The production mechanism for PBHs in general is not well understood, and for the
PEMNSs we shall make the simplifying assumption that they are first formed
when the accelerated expansion begins at $t=t_{DE}\sim 9.8$ Gy. For the expansion before $t_{DE}$ we shall assume
that the $\Lambda CDM$ model is approximately accurate.

\bigskip

\noindent
The subsequent expansion in the charged dark matter EAU model will in the future
depart markedly from the $\Lambda CDM$ case. We can regard this as advantageous
because the future fate of the universe in the conventional picture does have certain
distasteful features in terms of the extroverse, as we briefly
review.

\bigskip

\noindent
In the $\Lambda CDM$  model the introverse, or what is also called the visible 
universe, coincides with the extroverse at $t=t_{DE} \sim 9.8$ Gy with the common
radius

\begin{equation}
R_{EV}(t_{DE}) = R_{IV}(t_{DE}) =  39 Gly
\label{tDE}
\end{equation}

\bigskip

\noindent
The introverse expansion is limited by the speed of light and its radius increases
from Eq. (\ref{tDE}) to 45 Gly at the present time $t=t_0$ and asymptotes to

\begin{equation}
R_{IV} (t \rightarrow \infty) = 58 Gly
\label{RIVasymp}
\end{equation}

\bigskip

\noindent
The extroverse expansion is exponential and superluminal. Its radius increases
from its\\value 39 Gly in Eq. (\ref{tDE}) to 52 Gly at the present time $t=t_0$ and grows without limit
so that after a trillion years it attains the \\
{\it extremely} large value

\begin{equation}
R_{EV} (t  = 1 Ty) = 9.7 \times 10^{32} Gly.
\label{REVtrillion}
\end{equation}

\bigskip

\noindent
This future for the $\Lambda CDM$ scenario seems distasteful because the
introverse becomes of ever decreasing, and eventually vanishing, significance,
relative to the extroverse. At $t=1$Ty, all other galaxies would have
exited from the visible universe so that cosmology would no longer be possible.

\bigskip

\noindent
We shall for simplicity assume that
the PEMNSs are all formed between $t=t_{DE} \sim 9.8$ Gy and
$t_0 \sim 13.8$Gy. The
Friedmann equation ignoring radiation, during this time window, is

\begin{equation}
\left( \frac{\dot{a}}{a} \right)^2= \frac{\Lambda(t)}{3} + \frac {8\pi G}{3} \rho_{matter}
\label{Friedmann}
\end{equation}
where $\Lambda(t)$ is the cosmological parameter generated by the Coulomb
repulsion between the PEMNSs.
From Eq.(\ref{Friedmann}), with $a(t_0) = 1$ and constant $\Lambda$ as usually
assumed in the $\Lambda$CDM model we
would predict that

\begin{equation}
a(t \rightarrow \infty) \sim exp \left( \sqrt{ \frac{\Lambda}{3}} (t-t_0) \right)
\label{exponential}
\end{equation}

\bigskip

\noindent
However, in the case of charged dark matter, with no dark energy, we must
re-write \\Eq, ({\ref{Friedmann}) as

\begin{equation}
\left( \frac{\dot{a}}{a} \right)^2= \frac {8\pi G}{3} \rho_{PEMNS}+ \frac {8\pi G}{3} \rho_{matter}
\label{FriedmannPrime}
\end{equation}
in which

\begin{equation}
\rho_{matter} (t)  = \frac{\rho_{matter} (t_0)}{a(t)^3}
\label{rhomatter}
\end{equation}
where matter includes both normal matter and the uncharged dark matter.

\bigskip

\noindent
Of special interest in the present discussion is the expected future behaviour
of the charged dark matter

\begin{equation}
\rho_{PEMNS} (t) = \frac{\rho_{PEMNS} (t_0)}{a(t)^3}
\label{rhocPEMBHs}
\end{equation}
so that comparison of Eq.(\ref{Friedmann}) and Eq.(\ref{FriedmannPrime}) suggests
that the cosmological parameter $\Lambda(t)$ is predicted to decrease from its present value.

\bigskip

\noindent
More specifically, we find
that asymptotically  the scale factor will behave as if matter-dominated
and the cosmological parameter will decrease at large future times as a power

\begin{equation}
a(t\rightarrow \infty) \sim t^{\frac{2}{3}} ~~~~~~ \Lambda(t \rightarrow \infty) \sim t^{-2}.
\label{scale}
\end{equation}

\bigskip

\noindent
so that a trillion years in the future $\Lambda(t)$ will have decreased
by some four orders of magnitude relative to $\Lambda(t_0)$.

\bigskip

\noindent
According to the $\Lambda CDM$ model, we live at a special time in
cosmic history because of the density coincidence between dark matter
and dark energy.
In the case of charged dark matter replacing dark energy, the present
era is treated differently because the striking accelerated expansion,
discovered in 1998, is a temporary phenomenon at around the present time.
Acceleration 
began about 4 Gy ago at $t_{DE}= 9.8Gy = t_0-4 Gy$.

\bigskip

\noindent
Let us  discuss the future time evolution of the introverse and
extroverse in the case of charged dark matter.  For the introverse,
nothing changes from the $\Lambda CDM$ case. After a trillion years, the introverse radius
will be at its asymptotic value $R_{IV}=58 Gly$, as stated in Eq.(\ref{RIVasymp}).
By contrast, the future for the extroverse
is very different for charged dark matter. 
WIth the growth $a(t) \propto t^{\frac{2}{3}}$ we find
that the radius of the extroverse at $t=1$ Ty is 

\begin{equation}
R_{EV}(t=1Ty) \sim 900 Gly
\label{REVnew}
\end{equation}

\noindent
to be compared with the corresponding huge value
$9.7 \times 10^{32}$ Gly predicted by the $\Lambda CDM$ model. 
This means that if there still exist humans
in the Solar System, or at least in the Milky Way, their view
of the distant universe will include many billions of galaxies.

\bigskip

\noindent
In the $\Lambda CDM$ case, a hypothetical observational cosmologist, one trillion years
in the future,
could observe only the Milky Way and objects such as the Magellanic 
Clouds which are gravitationally bound to it, so that cosmology could become
an extinct science.
In the case of charged dark matter, for comparison, the time dependence
will allow about 180 billion out of a present trillion galaxies to
remain observable at $t=1Ty$ so that the view of the 
universe at that distant future time will look quite similar to 
the view at the present and will provide
 a conducive environment for cosmology, a trillion years in the
future.

\bigskip

\noindent
\section{Discussion}

\bigskip

\noindent
Admittedly we have studied only a simplified model of the visible universe
but it is sufficiently
realistic to take away two lessons:
\noindent
(1) Dark energy does not exist as a separate entity. It was a misidentification
of one part of dark matter which is composed of electrically-charged
extremely-massive primordial black holes.
\noindent
(2) The observed magnitude of the cosmological constant $|\Lambda| \sim (meV)^4$ is not surprisingly
small but is closely equal to the mean mass density of the universe. Distinguished authors have
provided clever explanations for an alleged smallness. They assume, however, that the theory of gravity beyond Einstein is based 
on quantum theories of fields or strings, although we do not yet
know this as an established fact.

\bigskip

\noindent
Our first argument for the existence of PEMNSs arose from approaching the holographic 
maximum entropy bound. This would be more compelling if the entropy 
of the universe were shown to be, as we suspect, a truly meaningful and useful concept. Regrettably 
this has not really happened, because it presently lacks a sufficiently rigorous underlying mathematical basis.

\bigskip

\noindent
Our second argument favouring PEMNSs is that they can replace
the notion of dark energy which was a
term inserted into
the Friedmann equation to parametrise the observed
accelerating expansion. Dark energy required repulsive gravity
which contradicted the assumption that gravity is always attractive.

\bigskip

\noindent
It seems more natural to ascribe the accelerated expansion to the other
long-range force, electromagnetism.
The same-sign charges of our PEMNSs can provide sufficient Coulomb 
electromagnetic repulsion. As we have shown, it can generate
a cosmological parameter $\Lambda(t_0)$ of the correct magnitude and the observed
equation of state with high accuracy.

\bigskip

\noindent
At first sight, our model appears counterintuitive because at the largest
length scales it seemed obvious to assume that gravitation
dominates. But this assumption need not be true in Nature. In our model we actually have
different regimes of length scales, and only in the one
characterising the Solar System, galaxies and clusters, does 
gravity dominate while, at larger, truly-cosmological scales, electromagnetism
takes over from gravity again.

\bigskip

\noindent
Based on our model and the precision of the wide-separation
expansion in $1/r$ we have predicted that the equation of state
associated with the cosmological constant to be
$\omega = -1$ to $O(10^{-14})$ accuracy and therefore,
since CMB perturbations are much larger,
this deviation from $\omega = -1$ is likely long
to remain unobservable
experimentally. If and when there emerges a convincing
(quantum?)gravity theory beyond Einstein, it could possibly
give further
corrections to the equation of state. We suspect, however, that such
corrections, if they exist, would be even smaller than the error
arising from the wide-separation approximation.

\end{document}